\documentstyle[preprint,aps,epsfig]{revtex}
\begin{document}
\draft
\title{Dielectric behaviors of corrugated membranes}
\author{K. W. Yu}
\address{Department of Physics, The Chinese University of Hong Kong, \\
         Shatin, New Territories, Hong Kong, China}
\maketitle

\begin{abstract}
We have employed our recently developed Green's function formalism to 
study the dielectric behavior of a model membrane, formed by two periodic 
interfaces separating two media of different dielectric constants.
The Maxwell's equations are converted into a surface integral equation; 
thus it greatly simplifies the solutions and yields accurate results 
for membranes of arbitrary shape.
The integral equation is solved and dielectric dispersion spectrum is 
obtained for a model corrugated membrane.
We report a giant dielectric dispersion as the amplitude of corrugation 
becomes large. 
\end{abstract}
\vskip 5mm

\section{Introduction}

It is known that lipid bilayers (abound in living cells membranes) 
exhibit a ripple phase of the bilayer-water interface, in a narrow 
temperature range \cite{smith}. 
The ripple phase is characterized by permanent wave-like deformations 
of the interfaces, the origin of which is still a widely debated topic. 
However, it was observed that a giant dielectric dispersion can occur 
in the radiofrequency range, signified by a large dielectric increment 
in that frequency range \cite{ripple}.
In practice, these membrane systems are under intensive research because 
they are responsible for delivery and retention of drugs \cite{bio-mimetic}.

We aim to obtain some results on the dielectric dispersion spectrum of 
corrugated membranes, by using our recently developed Green's function 
formalism of periodic interfaces \cite{Yu}.
In that formalism, we obtained an analytic expression for the Greenian 
that includes the effect of periodicity. 
In this work, we extend the Green's function formalism to compute
the local field distribution for a lipid bilayer membrane of arbitrary 
shape, separating two media of different dielectric constants. 
We will calculate the effective dielectric constant of the membrane 
subject to an ac applied field.

\section{Green's Function Formalism}

The Green's function formalism has been published recently \cite{Yu}. 
Here we re-iterate the formalism to establish notation. We will apply the 
formalism to a single interface and then extend to a bilayer membrane.
The electrostatic potential satisfies the Laplace's equation:
\begin{eqnarray}
\nabla \cdot [\epsilon({\bf r}) \nabla \phi({\bf r})] = -4\pi \rho({\bf r}),
\label{electrostatics}
\end{eqnarray}
with standard boundary conditions on the interface, where $\rho({\bf r})$
is the free charge density, $\epsilon({\bf r})$ equals $\epsilon_2$ in the 
host and $\epsilon_1$ in the embedding medium.

Let $V_1$ and $V_2$ be the volume of the embedding and host medium, 
separated by an interface $S$.
Denoting $\theta({\bf r})=1$ if ${\bf r} \in V_1$ and 0 otherwise,
leads to an integral equation \cite{Yu2000}:
\begin{eqnarray}
[1-u\theta({\bf r})]\phi({\bf r})
  =\phi_0({\bf r})  
    + {u\over 4\pi} \oint_S dS'\left[ \hat{\bf n}'
      \cdot \nabla' G({\bf r}, {\bf r}') \right] \phi({\bf r}'),
\label{integral}
\end{eqnarray}
where $u=1-\epsilon_1/\epsilon_2$, $\hat{\bf n}'$ is unit normal to $S$, 
$G({\bf r}, {\bf r}')=1/|{\bf r}-{\bf r}'|$ and 
$\phi_0$ is the solution of $\nabla^2 \phi_0=-4\pi \rho/\epsilon_2$.

Accordingly, our approach aims to solve a surface integral equation for
the potential at the expense of a two-step solution \cite{Yu2000}:
\begin{enumerate}
\item step 1: determine $\phi({\bf r})$ for all {\bf r} $\in S$
  by solving Eq.(\ref{integral}), and then
\item step 2: obtain $\phi({\bf r})$ for all {\bf r}
  by using Eq.(\ref{integral}) and the results of step 1.
\end{enumerate}
In step 1, we encounter a singularity when the integration variable
${\bf r}'$ approaches the point of observation {\bf r}.
To circumvent the problem, we take an infinitesimal volume around {\bf r}
and perform the surface integral analytically, we find \cite{Yu2000}
\begin{eqnarray}
\left( 1-{u\over 2}\right) \phi({\bf r})
  =\phi_0({\bf r}) 
    + {u\over 4\pi} \oint'_S
      dS'\left[ \hat{\bf n}' \cdot \nabla' G({\bf r}, {\bf r}') \right]
        \phi({\bf r}'), \ \ \ {\bf r} \in S,
\label{r_in_S}
\end{eqnarray}
where ``prime'' denotes a restricted integration which excludes
${\bf r}'={\bf r}$.
The (surface) integral equation (\ref{r_in_S}) can be solved for
$\phi({\bf r}\in S)$.

\section{Application to a periodic interface}

Here we apply the integral equation formalism to a periodic interface. 
Suppose the interface profile depends only on $x$, described by $y=f(x)$, 
where $f(x)$ is a periodic function of $x$ with period $L$: $f(x+L)=f(x)$. 
Without loss of generality, we will let $L=1$ in subsequent studies.
Thus medium 1 occupies the space $y<f(x)$ while medium 2 occupies 
the space $y>f(x)$ separated by the interface at $y=f(x)$.
The external field is ${\bf E}_0$ and 
$\phi_0({\bf r})=-{\bf E}_0 \cdot {\bf r}$ is the potential.
For a periodic system, $\phi({\bf r})$ is a periodic function of the 
lattice vector ${\bf T}$. In what follows, we adopt similar treatment
as the Korringa, Kohn and Rostoker (KKR) method \cite{KKR}
and rewrite the integral equation as:
\begin{eqnarray}
\left( 1-{u\over 2}\right) \phi({\bf r})
  =-{\bf E}_0 \cdot {\bf r} 
    + {u\over 4\pi} \oint'_S
      dS' \tilde{G}({\bf r}, {\bf r}') \phi({\bf r}'),
\label{unit-cell}
\end{eqnarray}
where the integration is performed within a {\em unit cell}. 
The structure Green's function (Greenian) is given by \cite{KKR}:
\begin{eqnarray}
\tilde{G}({\bf r}, {\bf r}')=\sum_{\bf T} \hat{\bf n}' \cdot \nabla' 
    G({\bf r}, {\bf r}'+{\bf T}) 
=\sum_{m} {2\hat{\bf n}' \cdot ({\bf r}-{\bf r}'-m\hat{\bf x})
  \over |{\bf r}-{\bf r}'-m\hat{\bf x}|^2 }.
\end{eqnarray}
We were able to evaluate the Greenian analytically \cite{Yu}:
\begin{eqnarray}
\tilde{G}(x,y;x',f(x')) 
  = {2\pi [f'(x') \sin 2\pi (x - x') - \sinh 2\pi (y - f(x'))] \over            
    \cos 2\pi (x - x') - \cosh 2\pi (y - f(x')) }.
\label{Greenian}
\end{eqnarray}
Eq.(\ref{Greenian}) is a truly remarkable result -- the analytic expression
is valid for an arbitrary interface profile.
If the point of observation $(x,y)$ is located at the interface, 
the Greenian has a finite limit as $x' \to x$:
\begin{eqnarray}
\tilde{G}(x,f(x);x',f(x')|x'\to x)={f''(x)\over 1+(f'(x))^2}.
\end{eqnarray}
We first solve Eq.(\ref{unit-cell}) for the potential $\phi(x,f(x))$ 
right at the interface:
\begin{eqnarray}
\left( 1-{u\over 2}\right) \phi(x,f(x))
  =-E_0 f(x) 
   + {u\over 4\pi} \int dx' \tilde{G}(x,f(x);x',f(x')) \phi(x',f(x')).
\end{eqnarray}
Then we use Eq.(\ref{integral}) to find the potential at any 
arbitrary point $(x,y) \notin S$, using the potential at the interface. 
\begin{eqnarray}
(1-u)\phi(x,y)=-E_0 y  
   + {u\over 4\pi} \int dx' \tilde{G}(x,y;x',f(x')) \phi(x',f(x')), \\
\phi(x,y)=-E_0 y  
   + {u\over 4\pi} \int dx' \tilde{G}(x,y;x',f(x')) \phi(x',f(x')),
\end{eqnarray}
for $(x,y) \in V_1$ and $(x,y) \notin V_1$ respectively.

\section{Application to corrugated membranes}

Here we extend the formalism to a bilayer membrane.
Consider two interface profiles described by $y=f_t(x)$, where $t=0,1$
denote the lower and upper interface profiles respectively. 
Again $f_t(x)$ is a periodic function of $x$. 
Thus medium 1 occupies the space $f_0(x)<y<f_1(x)$ while medium 2 occupies 
the space $y<f_0(x)$ and $y>f_1(x)$.
For the upper (lower) profile, $\hat{\bf n} \cdot \hat{\bf y} > 0$
($\hat{\bf n} \cdot \hat{\bf y} < 0$), thus the Greenian becomes
\begin{eqnarray}
\tilde{G}_{tt'}(x,f_t(x);x',f_{t'}(x')) 
= {2\pi (-1)^{t'} [f_{t'}'(x')\sin2\pi(x - x') 
  - \sinh2\pi(f_t(x) - f_{t'}(x'))] \over 
  \cos 2\pi(x - x') - \cosh2\pi(f_t(x) - f_{t'}(x'))}.
\end{eqnarray}
The effective dielectric constant $\epsilon_e$ of the bilayer membrane 
satisfies the relation:
$$
\epsilon_e E_0 V = \int_V \epsilon({\bf r}) E_y dV = 
\epsilon_2 E_0 V - u \epsilon_2 \int_{V_1} E_y dV,
$$
where $V_1$ is the volume of the embedded medium.
As ${\bf E}=-\nabla \phi$, the volume integration can be converted into
a surface integration by the Green's theorem. Moreover, for the upper 
(lower) profile, $\hat{\bf n} \cdot \hat{\bf y} > 0$ 
($\hat{\bf n} \cdot \hat{\bf y} < 0$), 
thus the effective dielectric constant becomes
\begin{eqnarray}
\epsilon_e  = \epsilon_2  - {u\epsilon_2 V_1\over E_0 V} \int_0^1 dx 
  \left[\phi(x,f_1(x)) - \phi(x,f_0(x)) \right].
\end{eqnarray}

To solve the integral equation, we express the potential at an 
arbitrary point into a mode expansion:
\begin{eqnarray}
\phi(x,y) = \sum_{jk} C_{jk} \psi_j(x) \xi_k(y-f_t(x)),
\label{mode}
\end{eqnarray}
where $\psi_j(x)$ and $\xi_k(y)$ are mode functions.
The potential on the interfaces suffices:
\begin{eqnarray}
\phi(x,f_t(x)) = \sum_{jk} C_{jk} \psi_j(x) \xi_k(0) 
 = \sum_{j} A_{j} \psi_j(x),
\label{mode-a}
\end{eqnarray}
where $A_j=\sum_k C_{jk} \xi_k(0)$. Here we make a few remarks on the 
choice of the mode functions. The choice of the mode function is somewhat 
arbitrary in theory. In practice, these functions should be simple and 
easy to use. Common choice ranges from extended mode functions 
like the Fourier series expansions to localized mode functions 
like the step and triangular functions \cite{Yu}.
Substituting the mode expansion Eq.(\ref{mode-a}) into Eq.(\ref{unit-cell}), 
the coefficients $A_i$ satisfy the matrix equation:
\begin{eqnarray}
\left[ \left( 1-{u\over 2}\right) {\sf B} - {u\over 4\pi} {\sf M} \right]
 {\bf A} = -E_0 {\bf V},
\end{eqnarray}
where 
\begin{eqnarray}
B_{ij} &=& \int dx \psi_i(x) \psi_j(x),
\\
M_{ij} &=& \int\int dxdx'\psi_i(x) \tilde{G}_{tt'}(x,f_t(x);x',f_{t'}(x')) 
  \psi_j(x'),
\\
V_{i} &=& \int dx \psi_i(x) f_t(x).
\end{eqnarray}
It should be remarked that the mode functions need not be orthonormal and
the matrix {\sf B} is non-diagonal in general.

\section{Numerical results}

As a model bilayer membrane, we adopt the interface profiles:
$$
f_t(x)=-a\left(\cos 2\pi x - 0.1 \sin 4\pi x \right) + (t-0.5),\ \ t=0,1,
$$
where $a$ is the amplitude of corrugation, and the sine function is added 
to upset the reflection symmetry about $x=0$ \cite{Yu}. 
The width of the bilayer membrane is thus unity.
We adopt the step functions for the mode expansions:
$$
\theta_i^h(x)=1,\ \ \ x_i-{h\over 2} < x < x_i+{h\over 2}\ \ \ {\rm and}\ \
\theta_i^h(x)=0,\ \ \ {\rm otherwise}, 
$$
where $h$ is the width of the step function. 
In what follows, we adopt 100 step functions both for the lower and upper 
profiles, equally spaced in the unit interval $x \in (-1/2,1/2]$. 
The integrals Eqs.(16)--(18) can be readily performed.

To study the dielectric behavior, we apply an ac field at a frequency 
$\omega$. The embedded medium inside the membrane has a complex dielectric 
constant 
$$
\tilde{\epsilon_1} = \epsilon_1 + {\sigma_1\over i\omega},
$$
where $\epsilon_1$ and $\sigma_1$ are the dielectric constant and 
conductivity of the embedded medium respectively, 
with $\omega$ being the frequency of the applied field. 
We adopt the following parameters in the calculation: 
$\epsilon_1=40, \sigma_1=160$, while $\epsilon_2=80$. Also let $V_1/V=1$.
The Maxwell-Wagner relaxation time of a planar interface is given by:
$$
\tau = \epsilon_1/\sigma_1.
$$
In Fig.\ref{Fig1}, we plot (a) the real and (b) imaginary parts of the 
complex effective dielectric constant $\epsilon_e$ normalized to 
$\epsilon_2$ as function of frequency for various amplitude of corrugation 
$a$ ranging from 0.1 to 1.0.  
As is evident from Fig.\ref{Fig1}, there is a giant dielectric dispersion 
as the amplitude of corrugation becomes large ($a \to 1.0$). 

In summary, we have employed the Green's function formalism to study the 
dielectric behavior of a corrugated membrane.
The integral equation is solved and the dielectric dispersion spectrum is 
obtained for a periodic corrugated membrane.
We should remark that the present formalism can readily be generalized 
to multi-layers systems as well as corrugations in two dimensions \cite{egg}.

\section*{Acknowledgments}
This work was supported by the Research Grants Council of the Hong Kong 
SAR Government under grant CUHK 4245/01P. K. W. Yu acknowledges useful
conversation with Professor Hong Sun.

\newpage

\begin{figure}[h]
\caption{Normalized (a) real and (b) imaginary parts of the complex 
effective dielectric constant of a corrugated membrane plotted as function 
of the frequency of applied field for various amplitudes of corrugation.
Note that a large dielectric dispersion occurs for large amplitudes 
of corrugation.} 
\label{Fig1}
\end{figure}

\newpage
\centerline{\epsfig{file=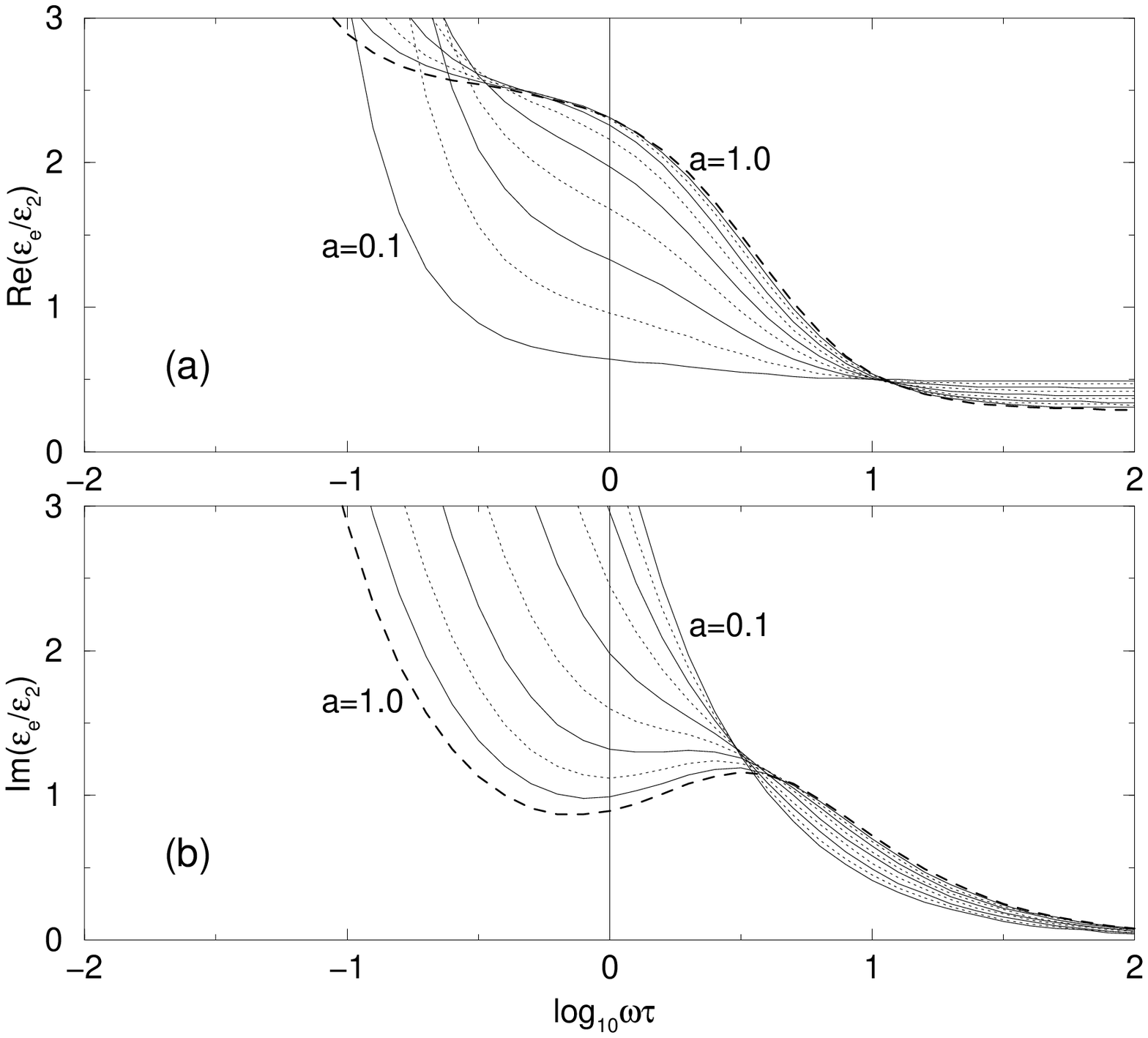,width=\linewidth}}
\centerline{Fig.1/Yu}

\end{document}